\documentclass[11pt]{article}

\usepackage{amsmath}

\newcommand{\N}{N\raise.7ex\hbox{\underline{$\circ $}}$\;$}

\begin{document}

\begin{center}

{\bf V.M. Red'kov, N.G. Tokarevskaya \\
Factorizations for 3-rotations   and polarization of the light \\
in Mueller-Stokes an Jones formalisms\\[2mm]
}
B.I. Stepanov Institute of Physics \\
National Academy of Sciences of Belarus  \\
68 Nezavisimosti ave., Minsk, 220072 BELARUS \\
redkov@dragon.bas-net.by

\end{center}

\begin{quotation}

Formulas describing all 2-element and 3-element factorizations of
arbitrary element of the groups SU(2) and SO(3,R) are  derived.
Six 2-element factorizations, $ (U_{2}U_{3}U'_{2}), \;
(U_{3}U_{2}U'_{3}), \; (U_{3}U_{1}U'_{3}), \; (U_{1}U_{3}U'_{1}),
\; (U_{1}U_{2}U'_{1}), \; (U_{2}U_{1}U'_{2})$, provide all
possible way to define Euler type angles; and  six 3-element ones,
$ (U_{1}U_{2}U_{3}), \; (U_{1}U_{3}U'_{2}),$ $ \;
(U_{2}U_{3}U_{1}), \; (U_{2}U_{1}U_{3}), \; (U_{3}U_{1}U_{2}), \;
(U_{3}U_{2}U_{1})$ provide  all possible ways to parameterize the
unitary and orthogonal groups by three elementary angles. In the
context the light polarization formalism of  Stokes-Mueller
vectors  and Jones spinors,
 relations  produced  give
 a base to resolve arbitrary pure polarization  rotators
 into all possible  sets of elementary rotators  of two or  three
 constituents.

\end{quotation}


\section{Introduction}

In this paper we will consider several problems naturally arising within the task of
finding different ways to parameterize the unitary SU(2) and rotation SO(3) groups
by  three angle variables. There exist two different classes of those. The first   based on
2-element factorizations:
\begin{eqnarray}
 (U_{2} U_{3} U'_{2}) \; , \qquad (U_{3}U_{2}U'_{3}) \; , \qquad (U_{3}U_{1}U'_{3})\; , \;
\nonumber
\\
(U_{1} U_{3} U'_{1}) \; , \qquad (U_{1} U_{2} U'_{1}) \; , \qquad
(U_{2}U_{1}U'_{2})\;,
\nonumber
\end{eqnarray}

\noindent
 provides all  possible ways to define Euler's angles.
The second  is based on 3-element factorizations:
\begin{eqnarray}
 (U_{1}U_{2}U_{3})\;, \qquad
(U_{1}U_{3}U'_{2})\;, \qquad
(U_{2}U_{3}U_{1})\;,
\nonumber
\\
(U_{2}U_{1}U_{3})\;, \qquad (U_{3}U_{1}U_{2})\;, \qquad
(U_{3}U_{2}U_{1})\;.
\nonumber
\end{eqnarray}

\noindent  In the literature, this second possibility  is used rarely, and
as a rule is only pointed out as existing one.

This rather abstract group theoretical  problems have  are of special  interest  in the context of
the light polarization formalism of  Stokes-Mueller vectors  and  Jones spinors.
Because  relations obtained     give a base to resolve arbitrary pure polarization  rotators
 into all possible  sets of elementary rotators  of two or  tree types.

 This paper is not a comprehensive  treatment of the physics of light polarization optics,
 it  only  discusses  group theoretical foundation for Stokes-Mueller vectors  and  Jones spinors
 formalisms. To fill up this  gap we give rather detailed bibliography  [1-110]
  on the subject,   though it  is not exhaustively complete.

\section*{2.  2-element  factorization, one special case}

In this Section we consider the following  2-element factorization:
\begin{eqnarray}
U = U_{1} U_{2} U'_{1}
= (x_{0} + ix_{1} \sigma_{1})  (y_{0} + iy_{2} \sigma_{2})
  (x'_{0} + ix'_{1} \sigma_{1})
\nonumber
\\
= n_{0} + i n_{1} \; \sigma_{1} + i n_{2} \;  \sigma_{2} +  i n_{3} \; \sigma_{3} \; ,
  \label{2.1}
\end{eqnarray}

\noindent which gives equations
\begin{eqnarray}
n_{0} =  y_{0}  \;(x_{0} x'_{0} - x_{1} x'_{1}) \; ,
\qquad
n_{1} = y_{0} \; ( x_{1} x'_{0} + x_{0} x'_{1})\; ,
\nonumber
\\
n_{2} = y_{2} \; ( x_{0} x'_{0} + x_{1} x'_{1})\; ,
\qquad
n_{3} = y_{2} \; ( -x_{1} x'_{0} + x_{0} x'_{1}) \; .
\label{2.2a}
\end{eqnarray}

\noindent We should resolve eqs. (\ref{2.2a}) under the variables $x_{0}, x_{1}; x'_{0}, x'_{1}; y_{0},  y_{2}$.
For  $y_{0},  y_{2}$ we have
\begin{eqnarray}
y_{0}= { n_{0} \over (x_{0} x'_{0} - x_{1} x'_{1})} = {n_{1}  \over ( x_{1} x'_{0} + x_{0} x'_{1}) }\; ,
\nonumber
\\
y_{2}  = { n_{2} \over  ( x_{0} x'_{0} + x_{1} x'_{1}) } =  {n_{3}  \over  ( -x_{1} x'_{0} + x_{0} x'_{1}) }\; .
\label{2.2b}
\end{eqnarray}

\noindent
In  eqs. (\ref{2.2a}),  one can exclude the variables $y_{0}$ and $y_{2}$:
\begin{eqnarray}
{n_{0} \over n_{1} } = {   (x_{0} x'_{0} - x_{1} x'_{1})   \over  ( x_{1} x'_{0} + x_{0} x'_{1}) } \; ,
\qquad
{n_{2} \over n_{3} } = { ( x_{0} x'_{0} + x_{1} x'_{1}) \over  ( -x_{1} x'_{0} + x_{0} x'_{1})  }
\nonumber
\end{eqnarray}

\noindent or
\begin{eqnarray}
n_{0} \; ( x_{1} x'_{0} + x_{0} x'_{1}) = n_{1} \; (x_{0} x'_{0} - x_{1} x'_{1}) \; ,
\nonumber
\\
n_{2} \; ( -x_{1} x'_{0} + x_{0} x'_{1}) = n_{3} \; ( x_{0} x'_{0} + x_{1} x'_{1}) \; .
\label{2.3}
\end{eqnarray}

Eqs. (\ref{2.3}) can be resolved as a linear system under the variables  $x_{0}, x_{1}$
\begin{eqnarray}
I\qquad
\left \{ \begin{array}{l}
x_{0}\; ( n_{0} x'_{1} - n_{1} x'_{0})  + x_{1} \; (n_{0} x'_{0} + n_{1} x'_{1}) = 0 \; , \\
x_{0}\; ( n_{2} x'_{1} - n_{3} x'_{0})  + x_{1} \; (-n_{2} x'_{0} - n_{3} x'_{1}) = 0 \; ,
\end{array} \right.
\label{2.4a}
\end{eqnarray}

\noindent or under the variables $x'_{0}, x'_{1}$:
\begin{eqnarray}
II\qquad
\left \{ \begin{array}{l}
x_{0}'\; ( n_{0} x_{1} - n_{1} x_{0})  + x'_{1} \; (n_{0} x_{0} + n_{1} x_{1}) = 0 \; ,
\\
x'_{0}\; ( -n_{2} x_{1} - n_{3} x_{0})  + x'_{1} \; (n_{2} x_{0} - n_{3} x_{1}) = 0 \; .
\end{array} \right.
\label{2.4b}
\end{eqnarray}

First, let us study eqs. (\ref{2.4a}):
\begin{eqnarray}
I \qquad
\mbox{det} \;
\left | \begin{array}{rr}
( n_{0} x'_{1} - n_{1} x'_{0})   &  (n_{0} x'_{0} + n_{1} x'_{1})   \\
( n_{2} x'_{1} - n_{3} x'_{0})   &  (-n_{2} x'_{0} - n_{3} x'_{1})
\end{array} \right | = 0 \; , \qquad \Longrightarrow
\nonumber
\\[2mm]
( n_{1} n_{3} -  n_{0}n_{2}) \; 2 x'_{0} x_{1} + (n_{0}n_{3}  + n_{1}n_{2}) (x_{0}^{'2} - x_{1}^{'2}) = 0 \; .
\label{2.5}
\end{eqnarray}

With the use of angle parametrization
$
x'_{0} =
 \cos {a ' \over 2} \;  , \; x'_{1} = \sin {a ' \over 2} \;,
$
 eq. (\ref{2.5}) takes the form
\begin{eqnarray}
( n_{1} n_{3} -  n_{0}n_{2}) \; \sin a'  + (n_{0}n_{3}  + n_{1}n_{2}) \;  \cos a'  = 0 \; ,
\nonumber
\end{eqnarray}

\noindent that is
\begin{eqnarray}
\mbox{tg}\; a' =  { n_{0}n_{3}  + n_{1}n_{2}  \over    n_{0}n_{2}  - n_{1} n_{3}  }\; ,
\label{2.7}
\end{eqnarray}

\noindent
and further
\begin{eqnarray}
\cos  a'  =  \sqrt{ {1 \over 1 + \mbox{tg}^{2} \; a'  }} =
  \pm
{ (n_{0} n_{2}  - n_{1} n_{3})   \over
 \sqrt{ n_{0}^{2}+n_{1} ^{2}}     \sqrt{n_{2}^{2}+n_{3}^{2} }    } \; ,
 \nonumber
 \\
 \sin a' = \cos a' \; \mbox{tg}\; a' =  \pm {  (n_{0} n_{3} + n_{1} n_{2} ) \over
\sqrt{ n_{0}^{2}+n_{1} ^{2}}     \sqrt{n_{2}^{2}+n_{3}^{2} }    } \; .
\label{2.9}
\end{eqnarray}

\noindent
Thus
\begin{eqnarray}
x'_{0} = \mu' \cos {a' \over 2} = \sqrt{ {1 + \cos a' \over 2}} \; , \;\; \mu ' = \pm 1 \; ,
\qquad
x'_{1} = \delta \sin {a' \over 2} = \sqrt{ {1- \cos a' \over 2}} \; , \;\; \delta ' = \pm 1 \; ,
\nonumber
\end{eqnarray}

\noindent or
\begin{eqnarray}
x'_{0} = \mu ' \; \sqrt{
  {1\over 2}  \pm { n_{0} n_{2}  - n_{1} n_{3}   \over 2
 \sqrt{ n_{0}^{2}+n_{1} ^{2} }     \sqrt{n_{2}^{2}+n_{3}^{2} }}} \; ,
\qquad
x'_{1} =  \delta' \; \sqrt{
  {1\over 2}  \mp        { n_{0} n_{2}  - n_{1} n_{3}   \over 2
 \sqrt{ n_{0}^{2}+n_{1} ^{2}}     \sqrt{n_{2}^{2}+n_{3}^{2} }}} \; .
\label{2.10}
\end{eqnarray}

\noindent
Turning to eqs. (\ref{2.4a}), we get two (equivalent) solutions:
\begin{eqnarray}
x_{0} =   \pm { (n_{0} x'_{0} + n_{1} x'_{1}) \over
\sqrt{( n_{0} x'_{1} - n_{1} x'_{0})^{2} +  (n_{0} x'_{0} + n_{1} x'_{1})^{2} }  } \; ,
\nonumber
\\
x_{1} = \pm  { ( n_{1} x'_{0}-n_{0} x'_{1} ) \over
\sqrt{( n_{0} x'_{1} - n_{1} x'_{0})^{2} +  (n_{0} x'_{0} + n_{1} x'_{1})^{2} }  } \; ;
\label{2.11a}
\end{eqnarray}
\begin{eqnarray}
x_{0} =   \pm { (n_{2} x'_{0} + n_{3} x'_{1}) \over
\sqrt{(  n_{2} x'_{1} - n_{3} x'_{0}  )^{2} +  ( -n_{2} x'_{0} - n_{3} x'_{1} )^{2} }  } \; ,
\nonumber
\\
x_{1} =  \pm  { (n_{2} x'_{1} - n_{3} x'_{0}) \over
\sqrt{(  n_{2} x'_{1} - n_{3} x'_{0}  )^{2} +  ( n_{2} x'_{0} +
n_{3} x'_{1} )^{2} }  }  \; .
\label{2.11b}
\end{eqnarray}

Now let us make the same with the system (\ref{2.4b}), which formally differs from (\ref{2.4a}) by evident changes
\begin{eqnarray}
n_{2} \longrightarrow - n_{2}\; , \qquad  x_{a} \longrightarrow x'_{a} \; ;
\nonumber
\end{eqnarray}

\noindent  from relationship
\begin{eqnarray}
II\qquad
\mbox{det}\;
\left | \begin{array}{rr}
( n_{0} x_{1} - n_{1} x_{0})   &   (n_{0} x_{0} + n_{1} x_{1}) \\
( -n_{2} x_{1} - n_{3} x_{0})  &   (n_{2} x_{0} - n_{3} x_{1})
\end{array} \right | = 0 \;  ,
\label{2.12}
\end{eqnarray}

\noindent
we get
\begin{eqnarray}
 \mbox{tg}\; a =  { - n_{0}n_{3}  + n_{1}n_{2}  \over    n_{0}n_{2}  + n_{1} n_{3}  }\; ,
\nonumber
\\
\cos  a  =  \pm \sqrt{ {1 \over 1 + \mbox{tg}^{2} \; a  }} =
  \pm
{ (n_{0} n_{2}  + n_{1} n_{3})   \over
 \sqrt{ n_{0}^{2}+n_{1} ^{2}}     \sqrt{n_{2}^{2}+n_{3}^{2} }    } \; ,
 \nonumber
 \\
 \sin a = \cos a \; \mbox{tg}\; a =  \pm {  (-n_{0} n_{3} + n_{1} n_{2} ) \over
\sqrt{ n_{0}^{2}+n_{1} ^{2}}     \sqrt{n_{2}^{2}+n_{3}^{2} }    } \; ;
\label{2.13}
\end{eqnarray}

\noindent also
\begin{eqnarray}
x_{0} = \cos {a \over 2} = \mu \sqrt{ {1 + \cos a \over 2}} \; , \;\; \mu  = \pm 1 \; ,
\nonumber
\\
x_{1} = \sin {a \over 2} = \delta \sqrt{ {1- \cos a \over 2}} \; ,  \;\; \delta  = \pm 1 \; ,
\\
x_{0} =  \mu  \; \sqrt{
  {1\over 2}  \pm { n_{0} n_{2}  + n_{1} n_{3}   \over 2
 \sqrt{ n_{0}^{2}+n_{1} ^{2} }     \sqrt{n_{2}^{2}+n_{3}^{2} }}} \; ,
 \nonumber
 \\
 x_{1} =  \delta \; \sqrt{
  {1\over 2}  \mp        {  n_{0} n_{2}  + n_{1} n_{3}   \over 2
 \sqrt{ n_{0}^{2}+n_{1} ^{2}}     \sqrt{n_{2}^{2}+n_{3}^{2} }}} \; .
\label{2.14}
\end{eqnarray}

\noindent
Turning to the system (\ref{2.4b}),
 we produce two (equivalent)  solutions:
\begin{eqnarray}
x_{0}' =  \pm { (n_{0} x_{0} + n_{1} x_{1}) \over
\sqrt{( n_{0} x_{1} - n_{1} x_{0}) ^{2}   + (n_{0} x_{0} + n_{1} x_{1})^{2}} }\; ,
\nonumber
\\
x_{1}' = \mp {   ( n_{0} x_{1} - n_{1} x_{0})    \over
\sqrt{( n_{0} x_{1} - n_{1} x_{0}) ^{2}   + (n_{0} x_{0} + n_{1} x_{1})^{2}} }\; ,
\label{2.14a}
\end{eqnarray}
\begin{eqnarray}
x_{0}' =  \pm { (n_{2} x_{0} - n_{3} x_{1}) \over  \sqrt{
( -n_{2} x_{1} - n_{3} x_{0})^{2} +   (n_{2} x_{0} - n_{3} x_{1})^{2}  }} \; ,
\nonumber
\\
x_{1}' =  \mp {
( n_{0} x_{1} - n_{1} x_{0})  \over  \sqrt{
( -n_{2} x_{1} - n_{3} x_{0})^{2} +   (n_{2} x_{0} - n_{3} x_{1})^{2}  }} \; .
\label{2.14b}
\end{eqnarray}

Evidently, the systems, I and II, are  equivalent,  so  they   must   provides   us
with the  same  solutions -- collect results together:

\vspace{3mm}
$I$
\begin{eqnarray}
 \cos  a'  =  \pm { (n_{0} n_{2}  - n_{1} n_{3})   \over
 \sqrt{ n_{0}^{2}+n_{1} ^{2}}     \sqrt{n_{2}^{2}+n_{3}^{2} }    } \; ,
\;
 \sin a' =  \pm {  (n_{0} n_{3} + n_{1} n_{2} ) \over
\sqrt{ n_{0}^{2}+n_{1} ^{2}}     \sqrt{n_{2}^{2}+n_{3}^{2} }    } \; ,
\nonumber
\end{eqnarray}
\begin{eqnarray}
x'_{0} =  \mu ' \; \sqrt{
  {1\over 2}  \pm { n_{0} n_{2}  - n_{1} n_{3}   \over 2
 \sqrt{ n_{0}^{2}+n_{1} ^{2} }     \sqrt{n_{2}^{2}+n_{3}^{2} }}} \; , \qquad
 x'_{1} =   \delta' \; \sqrt{
  {1\over 2}  \mp        { n_{0} n_{2}  - n_{1} n_{3}   \over 2
 \sqrt{ n_{0}^{2}+n_{1} ^{2}}     \sqrt{n_{2}^{2}+n_{3}^{2} }}} \; ,
\nonumber
\end{eqnarray}
\begin{eqnarray}
x_{0} =   \pm { (n_{0} x'_{0} + n_{1} x'_{1}) \over
\sqrt{   n_{0}^{2} + n_{1}^{2}    }  } \; ,
\qquad  x_{1} =  \pm  { ( n_{1} x'_{0}-n_{0} x'_{1} ) \over
\sqrt{  n_{0}^{2} + n_{1}^{2} }  } \; ;
\;\;
\nonumber
\\
x_{0} =  \pm { (n_{2} x'_{0} + n_{3} x'_{1}) \over
\sqrt{  n_{2}^{2} + n_{3}^{2} }  } \; ,
\qquad
x_{1} = \pm  { (n_{2} x'_{1} - n_{3} x'_{0}) \over
\sqrt{  n_{2}^{2} + n_{3}^{2} }  }  \; ;
\label{2.15}
\end{eqnarray}

$II
$
\begin{eqnarray}
\cos  a  =   \pm
{ (n_{0} n_{2}  + n_{1} n_{3})   \over
 \sqrt{ n_{0}^{2}+n_{1} ^{2}}     \sqrt{n_{2}^{2}+n_{3}^{2} }    } \; ,\qquad
 \sin a =
   \pm {  (-n_{0} n_{3} + n_{1} n_{2} ) \over
\sqrt{ n_{0}^{2}+n_{1} ^{2}}     \sqrt{n_{2}^{2}+n_{3}^{2} }    } \;,
\nonumber
\\
x_{0} =  \mu  \; \sqrt{
  {1\over 2}  \pm { n_{0} n_{2}  + n_{1} n_{3}   \over 2
 \sqrt{ n_{0}^{2}+n_{1} ^{2} }     \sqrt{n_{2}^{2}+n_{3}^{2} }}} \; ,
 \qquad
 x_{1} =  \delta \; \sqrt{
  {1\over 2}  \mp        {  n_{0} n_{2}  + n_{1} n_{3}   \over 2
 \sqrt{ n_{0}^{2}+n_{1} ^{2}}     \sqrt{n_{2}^{2}+n_{3}^{2} }}} \; ,
 \nonumber
 \\
x_{0}' = \pm { (n_{0} x_{0} + n_{1} x_{1}) \over
\sqrt{ n_{0}^{2} + n_{1}^{2}} }\; , \;\;
x_{1}' =  \mp {   ( n_{0} x_{1} - n_{1} x_{0})    \over
\sqrt{ n_{0}^{2} + n_{1}^{2}} }\; ,
\nonumber
\\
x_{0}' = \pm { (n_{2} x_{0} - n_{3} x_{1}) \over  \sqrt{
n_{2}^{2} + n_{3}^{2}  }} \; ,
\qquad
x_{1}' = \mp {
( n_{0} x_{1} - n_{1} x_{0})  \over  \sqrt{
n_{2}^{2} + n_{3}^{2}  }} \; .
\label{2.16}
\end{eqnarray}

\noindent
Let us use the most simple forms  (also one may omit $\pm$ )

\vspace{3mm}
for $a,a'$:

\begin{eqnarray}
 \cos  a'  =   { (n_{0} n_{2}  - n_{1} n_{3})   \over
 \sqrt{ n_{0}^{2}+n_{1} ^{2}}     \sqrt{n_{2}^{2}+n_{3}^{2} }    } \; ,
\quad
 \sin a' =   {  (n_{0} n_{3} + n_{1} n_{2} ) \over
\sqrt{ n_{0}^{2}+n_{1} ^{2}}     \sqrt{n_{2}^{2}+n_{3}^{2} }    } \;  ;
\nonumber
\\
\cos  a  =
{ (n_{0} n_{2}  + n_{1} n_{3})   \over
 \sqrt{ n_{0}^{2}+n_{1} ^{2}}     \sqrt{n_{2}^{2}+n_{3}^{2} }    } \; ,\quad
 \sin a =
    {  (-n_{0} n_{3} + n_{1} n_{2} ) \over
\sqrt{ n_{0}^{2}+n_{1} ^{2}}     \sqrt{n_{2}^{2}+n_{3}^{2} }    } \; ;
\label{2.17}
\end{eqnarray}

\noindent
and for ${a\over 2}, {a'\over 2}$:

\begin{eqnarray}
x'_{0} =  \cos{a'\over 2}=  \mu ' \; \sqrt{
  {1\over 2}  + { n_{0} n_{2}  - n_{1} n_{3}   \over 2
 \sqrt{ n_{0}^{2}+n_{1} ^{2} }     \sqrt{n_{2}^{2}+n_{3}^{2} }}} \; , \quad
 x'_{1} = \sin {a'\over 2} =  \delta' \; \sqrt{
  {1\over 2}  -        { n_{0} n_{2}  - n_{1} n_{3}   \over 2
 \sqrt{ n_{0}^{2}+n_{1} ^{2}}     \sqrt{n_{2}^{2}+n_{3}^{2} }}} \; ;
\nonumber
\\
x_{0} =  \cos{a\over 2}  = \mu  \; \sqrt{
  {1\over 2}  + { n_{0} n_{2}  + n_{1} n_{3}   \over 2
 \sqrt{ n_{0}^{2}+n_{1} ^{2} }     \sqrt{n_{2}^{2}+n_{3}^{2} }}} \; ,
 \qquad
 x_{1} =  \sin{a\over 2} = \delta \; \sqrt{
  {1\over 2}  -        {  n_{0} n_{2}  + n_{1} n_{3}   \over 2
 \sqrt{ n_{0}^{2}+n_{1} ^{2}}     \sqrt{n_{2}^{2}+n_{3}^{2} }}} \; ;
 \nonumber
 \\
 \label{2.18}
 \end{eqnarray}

Turning to eqs. (\ref{2.2b}), we should calculate expressions for
\begin{eqnarray}
y_{0}= { n_{0} \over (x_{0} x'_{0} - x_{1} x'_{1})} = {n_{1}  \over ( x_{1} x'_{0} + x_{0} x'_{1}) }\; ,
\nonumber
\\
y_{2}  = { n_{2} \over  ( x_{0} x'_{0} + x_{1} x'_{1}) } =  {n_{3}  \over  ( -x_{1} x'_{0} + x_{0} x'_{1}) }\; .
\label{2.19}
\end{eqnarray}

Let us introduce  angle variable
$y_{0} = \cos {b \over 2} \; , \; y_{2} = \sin {b \over 2}
$
and calculate $\sin b$ and $\cos b$; for definiteness let us use second expressions in (\ref{2.19}):
\begin{eqnarray}
\sin b = 2y_{0}y_{2} = 2 \; {n_{1} n_{3} \over
x_{1}^{2} (-1 + x^{'2}_{1} ) + x_{0}^{2} x_{1}^{'2} } =  2\; {n_{1} n_{3} \over x_{1}^{'2} - x_{1}^{2} } \; ;
\nonumber
\end{eqnarray}

\noindent
allowing for identity
\begin{eqnarray}
x_{1}^{'2} - x_{1}^{2} =
({1\over 2}  -        { n_{0} n_{2}  - n_{1} n_{3}   \over 2
 \sqrt{ n_{0}^{2}+n_{1} ^{2}}     \sqrt{n_{2}^{2}+n_{3}^{2} }} ) -
({1\over 2}  -        {  n_{0} n_{2}  + n_{1} n_{3}   \over 2
 \sqrt{ n_{0}^{2}+n_{1} ^{2}}     \sqrt{n_{2}^{2}+n_{3}^{2} }}) =
  {n_{1} n_{3} \over \sqrt{ n_{0}^{2}+n_{1} ^{2}}     \sqrt{n_{2}^{2}+n_{3}^{2} }}
\nonumber
\end{eqnarray}

\noindent we arrive at
\begin{eqnarray}
\sin b =  2  \; \sqrt{ n_{0}^{2}+n_{1} ^{2}}   \;   \sqrt{n_{2}^{2}+n_{3}^{2} } \; .
\label{2.20}
\end{eqnarray}

\noindent
It is readily verified that the result will be the same if one uses first expressions in (\ref{2.19}).
Let us calculate $\cos b$:
\begin{eqnarray}
\cos^{2} b = 1 - \sin^{2}b = (n_{0}^{2} +  n_{1}^{2})  + (n_{2}^{2} + n_{3}^{2}) -
4 (n_{0}^{2}+n_{1} ^{2} )    \;   (n_{2}^{2}+n_{3}^{2} ) =
(n_{0}^{2} +  n_{1}^{2} - n_{2}^{2} - n_{3}^{2})^{2}\; ,
\nonumber
\end{eqnarray}

\noindent so that
\begin{eqnarray}
\cos b = (n_{0}^{2} +  n_{1}^{2} - n_{2}^{2} - n_{3}^{2})\; .
\label{2.21}
\end{eqnarray}

Now let us calculate $y_{0}$ and $y_{2}$ from (\ref{2.19}).
Allowing for
\begin{eqnarray}
(x_{0} x'_{0}  \pm   x_{1} x'_{1}) = {1 \over 2 \sqrt{n_{0}^{2} + n_{1}^{2}} \; \sqrt{n_{2}^{2} + n_{3}^{2}}} \times
\nonumber
\\
\times   [ \; (\sqrt{n_{0}^{2} + n_{1}^{2}} \sqrt{ n_{2}^{2} + n_{3}^{2} } + (n_{0}n_{2}+n_{1}n_{3}) )^{1/2}\;
(\sqrt{n_{0}^{2} + n_{1}^{2}} \sqrt{ n_{2}^{2} + n_{3}^{2} } + (n_{0}n_{2}-n_{1}n_{3}) )^{1/2} \pm
\nonumber
\\
\pm  \; (\sqrt{n_{0}^{2} + n_{1}^{2}} \sqrt{ n_{2}^{2} + n_{3}^{2} } - (n_{0}n_{2}+n_{1}n_{3}) )^{1/2}\;
(\sqrt{n_{0}^{2} + n_{1}^{2}} \sqrt{ n_{2}^{2} + n_{3}^{2} } - (n_{0}n_{2}-n_{1}n_{3}) ) ^{1/2} \; ]
\nonumber
\\
= {1 \over 2 \sqrt{n_{0}^{2} + n_{1}^{2} }\; \sqrt{ n_{2}^{2} + n_{3}^{2}} } \;
[ \; (2n_{0}^{2}n_{2}^{2} + n_{0}^{2}n_{3}^{2} +n_{1}^{2} n_{2}^{2} +
2n_{0}n_{2} \sqrt{n_{0}^{2} + n_{1}^{2}} \sqrt{ n_{2}^{2} + n_{3}^{2} } )^{1/2}
\pm
\nonumber
\\
\pm (2n_{0}^{2}n_{2}^{2} + n_{0}^{2}n_{3}^{2} +n_{1}^{2} n_{2}^{2} -
2n_{0}n_{2} \sqrt{n_{0}^{2} + n_{1}^{2}} \sqrt{ n_{2}^{2} + n_{3}^{2} } )^{1/2}\; ]\; ,
\nonumber
\end{eqnarray}

\noindent
so that
\begin{eqnarray}
(x_{0} x'_{0}  +  x_{1} x'_{1}) = {
( n_{0}\;  \sqrt{ n_{2}^{2} + n_{3}^{2} }  + n_{2} \; \sqrt{n_{0}^{2} + n_{1}^{2} }  \; ) -
(n_{0}\;  \sqrt{ n_{2}^{2} + n_{3}^{2} }  - n_{2} \; \sqrt{n_{0}^{2} + n_{1}^{2}  } \; ) \over
2 \sqrt{n_{0}^{2} + n_{1}^{2}} \;  \sqrt{ n_{2}^{2} + n_{3}^{2} }} = {n_{2} \over   \sqrt{ n_{2}^{2} + n_{3}^{2} } }\; ,
\nonumber
\\
(x_{0} x'_{0}  -  x_{1} x'_{1}) = {
( n_{0}\;  \sqrt{ n_{2}^{2} + n_{3}^{2} }  + n_{2} \; \sqrt{n_{0}^{2} + n_{1}^{2} } \; ) +
(n_{0}\;  \sqrt{ n_{2}^{2} + n_{3}^{2} }  - n_{2} \; \sqrt{n_{0}^{2} + n_{1}^{2}  } \; ) \over
2 \sqrt{n_{0}^{2} + n_{1}^{2}} \; \sqrt{ n_{2}^{2} + n_{3}^{2}} } = {n_{0} \over   \sqrt{ n_{0}^{2} + n_{1}^{2} } } \; ,
\nonumber
\end{eqnarray}

\noindent thus
\begin{eqnarray}
y_{0} = { n_{0} \over (x_{0} x'_{0} - x_{1} x'_{1})}  =  \sqrt{ n_{2}^{2} + n_{3}^{2}}\; , \qquad
y_{2} = { n_{2} \over  ( x_{0} x'_{0} + x_{1} x'_{1}) } =  \sqrt{n_{0}^{2} + n_{1}^{2}} \; .
\label{2.22}
\end{eqnarray}

\section{All six types of 2-element factorizations}

There exist six  types of 2-element factorization:
\begin{eqnarray}
(1)\qquad U_{2}\;  U_{3} \; U_{2}' \; ,\qquad  (1)' \qquad U_{3}  \; U_{2} \; U_{3}' \; ;
\nonumber
\\[2mm]
(2)\qquad  U_{3}\;  U_{1} \; U_{3}' \; , \qquad  (2)' \qquad U_{1}  \; U_{3} \; U_{1}' \; ;
\nonumber
\\[2mm]
(3)\qquad  \underline{U_{1}\;  U_{2} \; U_{1}'} \; , \qquad  (3)' \qquad U_{2}  \; U_{1} \; U_{2}' \; .
\label{3.1}
\end{eqnarray}

\noindent in the previous Section we considered one (underlines) factorization $\underline{U_{1}\;  U_{2} \; U_{1}'}$.
We are to extend the above analysis to all six variants:
\begin{eqnarray}
U = n_{0} + i n_{1} \; \sigma_{1} + i n_{2} \;  \sigma_{2} +  i n_{3} \; \sigma_{3} \; ,
\nonumber
\end{eqnarray}
\begin{eqnarray}
(1) \qquad
U = U_{2} U_{3} U'_{2}
= (y_{0} + iy_{2} \sigma_{2})  (z_{0} + iz_{3} \sigma_{3})
  (y'_{0} + iy'_{2} \sigma_{2}) \; ,
  \nonumber
  \\
n_{0} =   z_{0}  \;(y_{0} y'_{0} - y_{2} y'_{2})  \; ,
\qquad
n_{3} =  z_{3} \; ( y_{0} y'_{0} + y_{2} y'_{2})\; ,
\nonumber
\\
n_{2} =   z_{0} \; ( y_{0} y'_{2} + y_{2} y'_{0}) \; ,
\qquad
n_{1} =   z_{3} \; ( y_{0} y'_{2} - y_{2} y'_{0}) \; .
\nonumber
\\[2mm]
(1)' \qquad
U = U_{3} U_{2} U'_{3}
= (z_{0} + iz_{3} \sigma_{3})  (y_{0} + iy_{2} \sigma_{2})
  (z'_{0} + iz'_{3} \sigma_{3}) \; ,
  \nonumber
  \\
n_{0} =   y_{0}  \;(z_{0} z'_{0} - z_{3} z'_{3})  \; ,
\qquad
n_{2} =  y_{2} \;  (z_{0} z'_{0} + z_{3} z'_{3})\; ,
\nonumber
\\
n_{3} =   y_{0} \; ( z_{0} z'_{3} + z_{3} z'_{0}) \; ,
\qquad
-n_{1} =   y_{2} \; (z_{0} z'_{3} - z_{3} z'_{0} ) \; ;
\label{3.2}
\end{eqnarray}

\begin{eqnarray}
(2) \qquad
U = U_{3} U_{1} U'_{3}
= (z_{0} + iz_{3} \sigma_{3})  (x_{0} + ix_{1} \sigma_{1})
  (z'_{0} + iz'_{3} \sigma_{3})\; ,
  \nonumber
  \\
 n_{0} =   x_{0}  \;(z_{0} z'_{0} - z_{3} z'_{3})  \; ,
\qquad
n_{1} =   x_{1} \; ( z_{0} z'_{0} + z_{3} z'_{3}) \; ,
\nonumber
\\
n_{3} =  x_{0} \; ( z_{0} z'_{3} + z_{3} z'_{0})\; , \qquad
n_{2} =   x_{1} \; ( z_{0} z'_{3} - z_{3} z'_{0}) \; ;
\nonumber
\\ [2mm]
(2)' \qquad
U = U_{1} U_{3} U'_{1}
= (x_{0} + ix_{1} \sigma_{1})  (z_{0} + iz_{3} \sigma_{3})
  (x'_{0} + ix'_{1} \sigma_{1})\; ,
  \nonumber
  \\
 n_{0} =   z_{0}  \;(x_{0} x'_{0} - x_{1} x'_{1})  \; ,
\qquad
n_{3} =   z_{3} \; ( x_{0} x'_{0} + x_{1} x'_{1}) \; ,
\nonumber
\\
n_{1} =  z_{0} \; ( x_{0} x'_{1} + x_{1} x'_{0})\; , \qquad
-n_{2} =   z_{3} \; ( x_{0} x'_{1} - x_{1} x'_{0}) \; ;
\label{3.3}
\end{eqnarray}

\begin{eqnarray}
(3) \qquad
U = U_{1} U_{2} U'_{1} = (x_{0} + ix_{1} \sigma_{1})  (y_{0} + iy_{2} \sigma_{2})
  (x'_{0} + ix'_{1} \sigma_{1})
\nonumber
\\
n_{0} =  y_{0}  \;(x_{0} x'_{0} - x_{1} x'_{1}) \; ,
\qquad
n_{2} = y_{2} \; (  x_{0} x'_{0} + x_{1} x'_{1}  )\; ,
\nonumber
\\
n_{1} = y_{0} \; (  x_{0} x'_{1}+ x_{1} x'_{0} )\; ,
\qquad
n_{3} = y_{2} \; (  x_{0} x'_{1} -x_{1} x'_{0} ) \; ;
\nonumber
\\[2mm]
(3)' \qquad
U = U_{2} U_{1} U'_{2} = (y_{0} + iy_{2} \sigma_{2})  (x_{0} + ix_{1} \sigma_{1})
(y'_{0} + iy'_{2} \sigma_{2})  \; ,
\nonumber
\\
n_{0} =    x_{0}  \;(y_{0} y'_{0} - y_{2} y'_{2})\; ,
\qquad
n_{1} = x_{1} \; ( y_{0} y'_{0} + y_{2} y'_{2}) \; ,
\nonumber
\\
n_{2} = x_{0} \; ( y_{0} y'_{2} + y_{2} y'_{0}) \; ,
\qquad
-n_{3} =  x_{1} \; ( y_{0} y'_{2} - y_{2} y'_{0})  \; .
\label{3.4}
\end{eqnarray}

All results obtained can be presented in the table:
\begin{eqnarray}
\left. \begin{array}{llll}
(1)\; \;\; - \;\; & (232)    &\qquad (y_{0},\;y_{2}) \;, \;  (z_{0},\;z_{3}) \;,\; (y'_{0},\;y'_{2}) \;,
\qquad &( n_{0},\; n_{2},\;n_{3},\;+n_{1}) \; ; \\
(1)' \;\; - \;\;  & (323)   &\qquad (z_{0},\;z_{3}) \;, \;  (y_{0},\;y_{2}) \;,\; (z'_{0},\;z'_{3}) \;,
\qquad &( n_{0},\; n_{3},\;n_{2},\;-n_{1}) \; ;\\[3mm]
(2) \;\;\; - \;\; &(313)  & \qquad (z_{0},\;z_{3}) \;, \;  (x_{0},\;x_{1}) \;,\; (z'_{0},\;z'_{3}) \;,
\qquad & ( n_{0},\; n_{3},\;n_{1},\;+n_{2}) \; ;\\
(2)' \;\; - \;\;  & (131) & \qquad  (x_{0},\;x_{1}) \;, \;  (z_{0},\;z_{3}) \;,\; (x'_{0},\;x'_{1}) \;,
\qquad & ( n_{0},\; n_{1},\;n_{3},\;-n_{2}) \; ; \\[3mm]
(3) \;\;\; - \;\; & (121)  & \qquad(x_{0},x_{1}) \;, \;  (y_{0},y_{2}) \;,\; (x'_{0},x'_{1}) \;,
\qquad & ( n_{0},\; n_{1},\; n_{2},\; +n_{3}) \; ;\\
(3)' \;\; - \;\; &(212)  & \qquad (y_{0},\; y_{2}) \;, \;  (x_{0},\; x_{1}) \;,\; (y'_{0},\; y'_{2}) \;,
\qquad &( n_{0},\; n_{2},\; n_{1},\; -n_{3}) \; .
\end{array} \right.
\label{3.5}
\end{eqnarray}

\noindent We see   that all six factorization according to (\ref{3.2}) -- (\ref{3.3}) -- (\ref{3.4}) have
 the same mathematical structure, therefore all six solutions can be produced by means of formal changes from
 results obtained for the case (3) -- (121) -- for simplicity we write down expressions for
 double angle variables:

\begin{eqnarray}
x_{0}=\cos {a\over 2}\;, \; x_{1}=\sin {a\over 2}\;,
\nonumber
\\
\cos  a  =
{ (n_{0} n_{2}  + n_{1} n_{3})   \over
 \sqrt{ n_{0}^{2}+n_{1} ^{2}}     \sqrt{n_{2}^{2}+n_{3}^{2} }    } \; ,\quad
 \sin a =
    {  (-n_{0} n_{3} + n_{1} n_{2} ) \over
\sqrt{ n_{0}^{2}+n_{1} ^{2}}     \sqrt{n_{2}^{2}+n_{3}^{2} }    } \; ;
 \nonumber
 \end{eqnarray}
 \begin{eqnarray}
y_{0}=\cos {b \over 2}\;, \; y_{2}=\sin {b\over 2}\;,
\nonumber
\\
\cos b = (n_{0}^{2} +  n_{1}^{2} - n_{2}^{2} - n_{3}^{2})\; ,
\quad
\sin b =  2  \; \sqrt{ n_{0}^{2}+n_{1} ^{2}}   \;   \sqrt{n_{2}^{2}+n_{3}^{2} } \; ;
\nonumber
\\
x'_{0}=\cos {a'\over 2}\;, \; x'_{1}=\sin {a'\over 2}\;,
\nonumber
\\
 \cos  a'  =   { (n_{0} n_{2}  - n_{1} n_{3})   \over
 \sqrt{ n_{0}^{2}+n_{1} ^{2}}     \sqrt{n_{2}^{2}+n_{3}^{2} }    } \; ,
\quad
 \sin a' =   {  (n_{0} n_{3} + n_{1} n_{2} ) \over
\sqrt{ n_{0}^{2}+n_{1} ^{2}}     \sqrt{n_{2}^{2}+n_{3}^{2} }    } \;  .
\nonumber
\\
\label{3.6}
\end{eqnarray}

\section{ 3-element factorization, special case}

For an arbitrary element from SU(2)
\begin{eqnarray}
U =  n_{0} + in_{1}\sigma_{1}
+ in_{2}\sigma_{2} + in_{3}\sigma_{3}, \qquad
n_{0}^{2} + n_{1}^{2} + n_{2}^{2} + n_{3}^{2} = +1
\nonumber
\end{eqnarray}

\noindent let us introduce a 3-element factorization
\begin{eqnarray}
U = U_{1} U_{2} U_{3} =
 (x_{0} +ix_{1}\sigma_{1})
(y_{0} +iy_{2}\sigma_{2})(z_{0} +iz_{3}\sigma_{3}) \; ;
\label{4.1}
\end{eqnarray}

\noindent which gives equations
\begin{eqnarray}
n_{0} = \; \; x_{0} y_{0} z_{0} +  x_{1} y_{2} z_{3} \; ,
\qquad
n_{1} = -x_{0} y_{2} z_{3} +  x_{1} y_{0} z_{0} \; ,
\nonumber
\\
n_{2} = \; \;x_{0} y_{2} z_{0} +  x_{1} y_{0} z_{3} \; ,
\qquad
n_{3} = \;\; x_{0} y_{0} z_{3} -  x_{1} y_{2} z_{0} \; .
\label{4.2}
\end{eqnarray}

\noindent
At given $n_{a}$ one should find $(x_{0},x_{1}),  (y_{0},y_{2}), (z_{0},z_{3})$, parameterized by angle variables
as follows
\begin{eqnarray}
x_{0} = \cos {a\over 2}\; , \; x_{1} = \sin {a\over 2} \; ,\quad
y_{0} = \cos {b\over 2}\; , \; y_{2} = \sin {b\over 2} \; , \quad
z_{0} = \cos {c\over 2}\; , \; z_{3} = \sin {c\over 2} \; .
\nonumber
\end{eqnarray}

Eqs. (\ref{4.2}) can be considered as two linear systems under $x_{0},x_{1}$:
\begin{eqnarray}
 \left \{  \begin{array}{r}
 x_{0} \; y_{0} z_{0} +  x_{1}\;  y_{2} z_{3} = n_{0}  \; , \\
  -x_{0} \; y_{2} z_{3} +  x_{1} \; y_{0} z_{0} = n_{1} \; ;
  \end{array} \right.
\qquad
 \left \{  \begin{array}{r}
x_{0}\;  y_{2} z_{0} +  x_{1} \; y_{0} z_{3} = n_{2}  \;  , \\
 x_{0} \; y_{0} z_{3} -  x_{1} \; y_{2} z_{0} = n_{3} \; .
  \end{array} \right.
\label{4.3a}
\end{eqnarray}

\noindent
Their solutions are respectively:
\begin{eqnarray}
x_{0} = { n_{0}\; y_{0}z_{0} - n_{1} \; y_{2} z_{3} \over  y^{2}_{0} z^{2}_{0} + y^{2}_{2} z^{2}_{3} } \; ,
\qquad
x_{1} = { n_{1}\; y_{0}z_{0} + n_{0} \; y_{2}z_{3} \over  y^{2}_{0}z^{2}_{0} + y^{2}_{2}z^{2}_{3}  }\; ;
\nonumber
\\
x_{0} = { n_{2}\; y_{2}z_{0} + n_{3} \; y_{0} z_{3} \over  y^{2}_{2} z^{2}_{0} + y^{2}_{0} z^{2}_{3} } \; ,
\qquad
x_{1} = {-  n_{3}\; y_{2}z_{0} + n_{2} \; y_{0} z_{3} \over  y^{2}_{2} z^{2}_{0} + y^{2}_{0} z^{2}_{3} } \; ,
\label{4.3b}
\end{eqnarray}

\noindent
One can exclude variables $ x_{0},x_{1}$ from eqs. (\ref{4.3b}):
\begin{eqnarray}
{ n_{0} y_{0}z_{0} - n_{1} \; y_{2} z_{3} \over  y^{2}_{0} z^{2}_{0} + y^{2}_{2} z^{2}_{3} } =
{ n_{2}\; y_{2}z_{0} + n_{3} \; y_{0} z_{3} \over  y^{2}_{2} z^{2}_{0} + y^{2}_{0} z^{2}_{3} } \; ,
\nonumber
\\
{ n_{1}\; y_{0}z_{0} + n_{0} \; y_{2}z_{3} \over  y^{2}_{0}z^{2}_{0} + y^{2}_{2}z^{2}_{3}  } =
{-  n_{3}\; y_{2}z_{0} + n_{2} \; y_{0} z_{3} \over  y^{2}_{2} z^{2}_{0} + y^{2}_{0} z^{2}_{3} }\; .
\label{4.3c}
\end{eqnarray}

Alternatively, eqs. (\ref{4.2}) can be considered as two linear systems under $y_{0},y_{2}$:
\begin{eqnarray}
 \left \{  \begin{array}{r}
 y_{0} \;  x_{0}  z_{0} +  y_{2} \; x_{1}   z_{3} = n_{0}  \; , \\
  y_{0}\;   x_{1}  z_{0}    -  y_{2} \; x_{0}  z_{3} = n_{1} \; ;
  \end{array} \right.
\qquad
 \left \{  \begin{array}{r}
 y_{0}\;   x_{1}  z_{3}  +   y_{2} \; x_{0}  z_{0} = n_{2}  \;  , \\
 y_{0}\;  x_{0}  z_{3} -  y_{2}\;  x_{1}  z_{0} = n_{3} \; .
  \end{array} \right.
\label{4.4a'}
\end{eqnarray}

\noindent
Their solutions are respectively:
\begin{eqnarray}
y_{0} = {n_{0} x_{0} + n_{1} x_{1} \over z_{0} } \;  , \qquad
y_{2} = {-n_{1} x_{0} + n_{0} x_{1} \over z_{3}} \; ;
\nonumber
\\
y_{0} = {n_{2} x_{1} + n_{3} x_{0} \over z_{3} } \;  , \qquad
y_{2} = {-n_{3} x_{1} + n_{2} x_{0} \over z_{0}} \; .
\label{4.4b'}
\end{eqnarray}

\noindent
Excluding the variables $y_{0},y_{2}$, we get
\begin{eqnarray}
{n_{0} x_{0} + n_{1} x_{1} \over z_{0} }= {n_{2} x_{1} + n_{3} x_{0} \over z_{3} } \; , \qquad
{-n_{1} x_{0} + n_{0} x_{1} \over z_{3}} = {-n_{3} x_{1} + n_{2} x_{0} \over z_{0}} \; .
\label{4.4c}
\end{eqnarray}

Alternatively, eqs. (\ref{4.2}) can be considered as two linear systems under $z_{0},z_{3}$:
\begin{eqnarray}
 \left \{  \begin{array}{r}
z_{0} \;  y_{0}   x_{0}   +  z_{3} \; y_{2}  x_{1}    = n_{0}  \; , \\
z_{0}  \;  y_{0}   x_{1}      - z_{3} \; y_{2}  x_{0}   = n_{1} \; ;
  \end{array} \right.
\qquad
 \left \{  \begin{array}{r}
z_{0} \;   y_{2}  x_{0}   +z_{3}\;  y_{0}   x_{1}     = n_{2}  \;  , \\
   -z_{0}\;  y_{2}  x_{1}    + z_{3} \; y_{0}  x_{0} = n_{3} \; .
  \end{array} \right.
\label{4.5a}
\end{eqnarray}

\noindent
Their solutions are respectively:
\begin{eqnarray}
z_{0} = {n_{0} x_{0} + n_{1} x_{1} \over y_{0} } \;  , \qquad
z_{3} = {-n_{1} x_{0} + n_{0} x_{1} \over y_{2}} \; ;
\nonumber
\\
z_{0} = {n_{2} x_{0} - n_{3} x_{1} \over y_{2} } \;  , \qquad
z_{3} = {n_{3} x_{0} + n_{2} x_{1} \over y_{0}} \; .
\label{4.5b}
\end{eqnarray}

\noindent
Excluding the variables $z_{0},z_{3}$, we get
\begin{eqnarray}
{n_{0} x_{0} + n_{1} x_{1} \over y_{0} } =
{n_{2} x_{0} - n_{3} x_{1} \over y_{2} }\; , \qquad
{-n_{1} x_{0} + n_{0} x_{1} \over y_{2}} =
{n_{3} x_{0} + n_{2} x_{1} \over y_{0}} \; .
\label{4.5c}
\end{eqnarray}

Two last variants, (\ref{4.4c}) and (\ref{4.5c}),  seem to be simpler than (\ref{4.3c}).
First, let us consider the variant (\ref{4.4c}):
\begin{eqnarray}
\left \{ \begin{array}{l}
z_{0} \; (n_{2} x_{1} + n_{3} x_{0}) \; - \; z_{3} \;(n_{0} x_{0} + n_{1} x_{1} ) = 0   \; , \\
 z_{0} (-n_{1} x_{0} + n_{0} x_{1} )  -   z_{3} (-n_{3} x_{1} + n_{2} x_{0}) = 0 \; .
\end{array} \right.
\label{4.6a}
\end{eqnarray}

\noindent From vanishing the determinant
\begin{eqnarray}
\left | \begin{array}{cc}
(n_{2} x_{1} + n_{3} x_{0})   & -  (n_{0} x_{0} + n_{1} x_{1} )  \\
(-n_{1} x_{0} + n_{0} x_{1} ) &  - (-n_{3} x_{1} + n_{2} x_{0})
\end{array} \right | =0
\nonumber
\end{eqnarray}

\noindent it follows
\begin{eqnarray}
n_{2} n_{3} \; x_{1}^{2} -n_{2}^{2}\; x_{0} x_{1} + n_{3}^{2}\; x_{0} x_{1} - n_{2} n_{3}\; x_{0}^{2}
-n_{0}n_{1}\; x_{0}^{2} +n_{0}^{2} \; x_{0} x_{1} - n_{1}^{2} \; x_{0}x_{1}  + n_{0}n_{1}\; x_{1}^{2} = 0 \; ,
\nonumber
\end{eqnarray}

\noindent which may be rewritten as
\begin{eqnarray}
- (n_{0}n_{1} + n_{2} n_{3}    )(x_{0}^{2} - x_{1}^{2} ) +
(n_{0}^{2}  + n_{3}^{2}  - n_{1}^{2} -n_{2}^{2}  ) \; x_{0} x_{1} = 0 \; ,
\nonumber
\\
\mbox{tg}\; a =   { 2n_{2} n_{3}  + 2n_{0} n_{1}
 \over  n_{0}^{2}  + n_{3}^{2}    - n_{1}^{2} -  n_{2}^{2}    } \; ; \qquad \qquad
\label{4.6b}
\end{eqnarray}

\noindent  expressions for $\cos a$ and $\sin a$ are
\begin{eqnarray}
\cos a = { n_{0}^{2}  + n_{3}^{2}    - n_{1}^{2} -  n_{2}^{2} \over
\sqrt {
(n_{0}^{2}  + n_{3}^{2}    - n_{1}^{2} -  n_{2}^{2})^{2} +  ( 2 n_{2} n_{3}  + 2n_{0} n_{1} )^{2}  } } \; ,
\nonumber
\\
\sin a = {  2  n_{2} n_{3}  + 2 n_{0} n_{1}  \over \sqrt{
(n_{0}^{2}  + n_{3}^{2}    - n_{1}^{2} -  n_{2}^{2} ) ^{2} +  ( 2 n_{2} n_{3}  + 2  n_{0} n_{1} )^{2}  }} \; .
\label{4.6c}
\end{eqnarray}

Now, in the same manner let us  consider the variant
(\ref{4.5c}):
\begin{eqnarray}
\left \{ \begin{array}{l}
 y_{0} \; (n_{2} x_{0} - n_{3} x_{1})  - y_{2} \; (n_{0} x_{0} + n_{1} x_{1} )  =0 \; ,\\
y_{0} (-n_{1} x_{0} + n_{0} x_{1})  -  y_{2} (n_{3} x_{0} + n_{2} x_{1}) = 0 \; .
\end{array} \right.
\label{4.7a}
\end{eqnarray}

\noindent
From vanishing the determinant
\begin{eqnarray}
\left | \begin{array}{l}
(n_{2} x_{0} - n_{3} x_{1}) -  (n_{0} x_{0} + n_{1} x_{1} ) \\
(-n_{1} x_{0} + n_{0} x_{1})  -  (n_{3} x_{0} + n_{2} x_{1})
\end{array} \right | = 0
\nonumber
\end{eqnarray}

\noindent it follows
\begin{eqnarray}
-n_{2} n_{3} \; x_{0}^{2}  -  n_{2}^{2} \; x_{0}x_{1} + n_{3}^{2} \; x_{0} x_{1} + n_{3}  n_{2} \; x_{1}^{2}
- n_{0} n_{1}\; x_{0}^{2} + n_{0}^{2}\; x_{0} x_{1} - n_{1}^{2} \; x_{0} x_{1} + n_{0} n_{1} \; x_{1}^{2} = 0 \; ,
\nonumber
\end{eqnarray}

\noindent or
\begin{eqnarray}
-(n_{2} n_{3}  + n_{0} n_{1}  )\; (  x_{0}^{2}  - x_{1}^{2} ) +
(n_{0}^{2}  + n_{3}^{2}    - n_{1}^{2} -  n_{2}^{2}   )  \; x_{0}x_{1} =0 \; ;
\nonumber
\end{eqnarray}

\noindent it may be written differently
\begin{eqnarray}
-(2n_{2} n_{3}  + 2n_{0} n_{1}  )\; \cos a  +
(n_{0}^{2}  + n_{3}^{2}    - n_{1}^{2} -  n_{2}^{2}   )  \; \sin a  =0 \; ,
\nonumber
\end{eqnarray}

\noindent  which coincides with (\ref{4.6b}) as it could be expected in advance.

Turning back to  (\ref{4.6a})
\begin{eqnarray}
\left \{ \begin{array}{l}
z_{0} (n_{2} x_{1} + n_{3} x_{0}) -  z_{3} (n_{0} x_{0} + n_{1} x_{1} ) = 0   \; , \\
 z_{0} (-n_{1} x_{0} + n_{0} x_{1} )  -   z_{3} (-n_{3} x_{1} + n_{2} x_{0}) = 0 \; .
\end{array} \right.
\nonumber
\end{eqnarray}

\noindent
we get  (second  equivalent variant is omitted)
\begin{eqnarray}
z_{0}= \cos {c \over 2} = { n_{0} x_{0} + n_{1} x_{1}   \over  \sqrt {
(n_{0} x_{0} + n_{1} x_{1} ) ^{2} + (n_{2} x_{1} + n_{3} x_{0})^{2}}} \; ,
\nonumber
\\
z_{3}= \sin {c \over 2}  = {  n_{2} x_{1} + n_{3} x_{0}  \over  \sqrt {
(n_{0} x_{0} + n_{1} x_{1} ) ^{2} + (n_{2} x_{1} + n_{3} x_{0})^{2}}}  \; .
\end{eqnarray}

\noindent
Let us calculate  $\sin c $:
\begin{eqnarray}
\sin c = 2z_{0}z_{3} =
{ (n_{0}n_{2} + n_{1}n_{3}) \; \sin a  + n_{0}n_{3}\;  (1 + \cos a) + n_{1}n_{2}\; (1 - \cos a)  \over
(n_{0} x_{0} + n_{1} x_{1} ) ^{2} + (n_{2} x_{1} + n_{3} x_{0})^{2}  }=
\nonumber
\\
=  {1 \over
(n_{0} x_{0} + n_{1} x_{1} ) ^{2} + (n_{2} x_{1} + n_{3} x_{0})^{2}  } \; \times \hspace{20mm}
\nonumber
\\
\times \; [  (n_{0}n_{3}+n_{1}n_{2}) +  { (n_{0} n_{2} + n_{1}n_{3}) (2  n_{2} n_{3}  + 2 n_{0} n_{1} )  +(n_{0}n_{3}-n_{1}n_{2})
(n_{0}^{2}  + n_{3}^{2}    - n_{1}^{2} -  n_{2}^{2}) \over
\sqrt {
(n_{0}^{2}  + n_{3}^{2}    - n_{1}^{2} -  n_{2}^{2})^{2} +  ( 2 n_{2} n_{3}  + 2n_{0} n_{1} )^{2}  } } \;
  ] =
\nonumber
\\
=  {(n_{0}n_{3}+n_{1}n_{2}) \over
(n_{0} x_{0} + n_{1} x_{1} ) ^{2} + (n_{2} x_{1} + n_{3} x_{0})^{2}  } \;\;
 [ \; 1 +  { 1 \over
\sqrt {
(n_{0}^{2}  + n_{3}^{2}    - n_{1}^{2} -  n_{2}^{2})^{2} +  ( 2 n_{2} n_{3}  + 2n_{0} n_{1} )^{2}  } } \;
  ] \; .
\nonumber
\end{eqnarray}

\noindent
Finding expression for  denominator
\begin{eqnarray}
(n_{0} x_{0} + n_{1} x_{1} ) ^{2} + (n_{2} x_{1} + n_{3} x_{0})^{2}  =
\nonumber
\\
=
(n_{0}^{2} + n_{3}^{2}) { 1+\cos a \over 2} + ( n_{1}^{2} +n_{2}^{2}){1- \cos a \over 2} +
(n_{0}n_{1}+n_{2}n_{3}) \sin a =
\nonumber
\\
={1\over 2} \; [ 1 + (n_{0}^{2} + n_{3}^{2} - n_{1}^{2} -n_{2}^{2})\cos a + (2n_{0}n_{1}+2n_{2}n_{3}) \sin a ]=
\nonumber
\\
{1 \over 2}\; [ \;1 +
\sqrt{ (n_{0}^{2}  + n_{3}^{2}    - n_{1}^{2} -  n_{2}^{2})^{2} +  ( 2 n_{2} n_{3}  + 2n_{0} n_{1} )^{2}  }\; ] \; ,
\end{eqnarray}

\noindent we arrive at
\begin{eqnarray}
\sin c = { 2(n_{0}n_{3}+n_{1}n_{2})  \over
\sqrt{ (n_{0}^{2}  + n_{3}^{2}    - n_{1}^{2} -  n_{2}^{2})^{2} +  ( 2 n_{2} n_{3}  + 2n_{0} n_{1} )^{2}  } } \; .
\label{4.8a}
\end{eqnarray}

\noindent
Now let us calculate $\cos c$:
\begin{eqnarray}
\cos c =  z_{0}^{2} - z_{3}^{2} =
{  (n_{0}^{2} - n_{3}^{2}) \; (1 + \cos a )  + ( n_{1}^{2} - n_{2}^{2})   (1 - \cos a )
 + 2(n_{0} n_{1}-n_{2} n_{3}) \sin a
 \over
2\;[  \; (n_{0} x_{0} + n_{1} x_{1} ) ^{2} + (n_{2} x_{1} + n_{3} x_{0})^{2}  \; ]} =
\nonumber
\\
=
[ \;1 +
\sqrt{ (n_{0}^{2}  + n_{3}^{2}    - n_{1}^{2} -  n_{2}^{2})^{2} +  ( 2 n_{2} n_{3}  + 2n_{0} n_{1} )^{2}  }\;\; ] \times
\nonumber
\\
\times
[\; (n_{0}^{2} - n_{3}^{2} + n_{1}^{2} - n_{2}^{2}) +
(n_{0}^{2} - n_{3}^{2} - n_{1}^{2} + n_{2}^{2}) \cos a + 2(n_{0} n_{1}-n_{2} n_{3}) \sin a  \; ] =
\nonumber
\\
\;[ \;1 +
\sqrt{ (n_{0}^{2}  + n_{3}^{2}    - n_{1}^{2} -  n_{2}^{2})^{2} +  ( 2 n_{2} n_{3}  + 2n_{0} n_{1} )^{2}  }\; ] \times
\nonumber
\\
\times \; (n_{0}^{2} - n_{3}^{2} + n_{1}^{2} - n_{2}^{2}) \; [\; 1 + {1 \over
\sqrt{ (n_{0}^{2}  + n_{3}^{2}    - n_{1}^{2} -  n_{2}^{2})^{2} +  ( 2 n_{2} n_{3}  + 2n_{0} n_{1} )^{2}  } }
 \; ] \; ,
\end{eqnarray}

\noindent
that is
\begin{eqnarray}
\cos c = {(n_{0}^{2} - n_{3}^{2} + n_{1}^{2} - n_{2}^{2})\over
\sqrt{ (n_{0}^{2}  + n_{3}^{2}    - n_{1}^{2} -  n_{2}^{2})^{2} +  ( 2 n_{2} n_{3}  + 2n_{0} n_{1} )^{2}  } }\; .
\label{4.8b}
\end{eqnarray}

\noindent
It is a matter of simple calculation to verify the identity
\begin{eqnarray}
(n_{0}^{2}  + n_{3}^{2}    - n_{1}^{2} -  n_{2}^{2})^{2} +  ( 2 n_{2} n_{3}  + 2n_{0} n_{1} )^{2}=
(n_{0}^{2} - n_{3}^{2} + n_{1}^{2} - n_{2}^{2})^{2} + (2n_{0}n_{3}+2n_{1}n_{2})^{2} \; .
\label{4.8c}
\end{eqnarray}

Thus, the angles $a$ and $c$ are determined by relations:
\begin{eqnarray}
\cos a = { n_{0}^{2}  + n_{3}^{2}    - n_{1}^{2} -  n_{2}^{2} \over
\sqrt {
(n_{0}^{2}  + n_{3}^{2}    - n_{1}^{2} -  n_{2}^{2})^{2} +  ( 2 n_{2} n_{3}  + 2n_{0} n_{1} )^{2}  } } \; ,
\nonumber
\\
\sin a = {  2  n_{2} n_{3}  + 2 n_{0} n_{1}  \over \sqrt{
(n_{0}^{2}  + n_{3}^{2}    - n_{1}^{2} -  n_{2}^{2} ) ^{2} +  ( 2 n_{2} n_{3}  + 2  n_{0} n_{1} )^{2}  }} \; ,
\nonumber
\\
\cos c = {(n_{0}^{2} - n_{3}^{2} + n_{1}^{2} - n_{2}^{2})\over
\sqrt{  (n_{0}^{2} - n_{3}^{2} + n_{1}^{2} - n_{2}^{2})^{2} + (2n_{0}n_{3}+2n_{1}n_{2})^{2} } } \; ,
\nonumber
\\
\sin c = { 2(n_{0}n_{3}+n_{1}n_{2})  \over
\sqrt{  (n_{0}^{2} - n_{3}^{2} + n_{1}^{2} - n_{2}^{2})^{2} + (2n_{0}n_{3}+2n_{1}n_{2})^{2}  } }\; .
\label{4.9}
\end{eqnarray}

Now, turning to (\ref{4.7a})
\begin{eqnarray}
\left \{ \begin{array}{l}
 y_{0} (n_{2} x_{0} - n_{3} x_{1}) -y_{2} (n_{0} x_{0} + n_{1} x_{1} )  =0 \; ,\\
y_{0} (-n_{1} x_{0} + n_{0} x_{1})  -  y_{2} (n_{3} x_{0} + n_{2} x_{1}) = 0 \; ,
\end{array} \right.
\end{eqnarray}

\noindent
we get expressions for $y_{0}, y_{2}$:
\begin{eqnarray}
y_{0}= {  (n_{0} x_{0} + n_{1} x_{1} )
\over \sqrt {
(n_{2} x_{0} - n_{3} x_{1})^{2} +  (n_{0} x_{0} + n_{1} x_{1} )^{2}  } } \; ,
\nonumber
\\
y_{2}= {  (n_{2} x_{0} - n_{3} x_{1})
\over \sqrt {
(n_{2} x_{0} - n_{3} x_{1})^{2} + (n_{0} x_{0} + n_{1} x_{1} )^{2}  }} \; .
\label{4.10a}
\end{eqnarray}

Let us calculate $\sin b$:
\begin{eqnarray}
\sin b= 2 y_{0}y_{2} = {n_{0}n_{2} (1+\cos a)  - n_{1}n_{3} (1- \cos a) +(  n_{1}n_{2} -n_{0}n_{3} ) \sin  a
\over
(n_{2} x_{0} - n_{3} x_{1})^{2} +  (n_{0} x_{0} + n_{1} x_{1} )^{2}   } \; .
\nonumber
\end{eqnarray}

\noindent
Using expression for numerator
\begin{eqnarray}
n_{0}n_{2} (1+\cos a)  - n_{1}n_{3} (1- \cos a) +(  n_{1}n_{2} -n_{0}n_{3} ) \sin  a
=
\nonumber
\\
=
( n_{0}n_{2} - n_{1}n_{3} ) +  ( n_{0}n_{2} + n_{1}n_{3} ) \cos a
+(  n_{1}n_{2} -n_{0}n_{3} ) \sin  a
=
\nonumber
\\
= ( n_{0}n_{2} - n_{1}n_{3} ) \; [\; 1 + { n_{0}^{2} -n_{3}^{2} +n_{1}^{2} -n_{2}^{2} \over
\sqrt {
(n_{0}^{2}  + n_{3}^{2}    - n_{1}^{2} -  n_{2}^{2})^{2} +  ( 2 n_{2} n_{3}  + 2n_{0} n_{1} )^{2}  } }\; ]
\nonumber
\end{eqnarray}

\noindent we get
\begin{eqnarray}
\sin b =
 { ( n_{0} n_{2} - n_{1}n_{3} )  \over
(n_{2} x_{0} - n_{3} x_{1})^{2} +  (n_{0} x_{0} + n_{1} x_{1} )^{2}   } \;
[\; 1 + { n_{0}^{2}  +n_{1}^{2} -n_{2}^{2}  -n_{3}^{2} \over
\sqrt {
(n_{0}^{2}  + n_{3}^{2}    - n_{1}^{2} -  n_{2}^{2})^{2} +  ( 2 n_{2} n_{3}  + 2n_{0} n_{1} )^{2}  } }\; ]\; .
\nonumber
\end{eqnarray}

\noindent Further, allowing for the expression for denominator
\begin{eqnarray}
(n_{2} x_{0} - n_{3} x_{1})^{2} +  (n_{0} x_{0} + n_{1} x_{1} )^{2}=
\nonumber
\\
= (n_{2}^{2} +n_{0}^{2} ) {1+ \cos a \over 2 } + (n_{3}^{2} + n_{1}^{2}) {1 - \cos a \over 2} +
(n_{0} n_{1} -n_{2}n_{3}) \sin a =
\nonumber
\\
={1 \over 2} \; [ 1 + ( n_{2}^{2} +n_{0}^{2}-n_{3}^{2} - n_{1}^{2}) \cos a  +
2(n_{0} n_{1} -n_{2}n_{3}) \sin a \; ] =
\nonumber
\\
= {1 \over 2} \; [ 1 +  { n_{0}^{2} + n_{1}^{2} -n_{2}^{2} -n_{3}^{2} \over
\sqrt {
(n_{0}^{2}  + n_{3}^{2}    - n_{1}^{2} -  n_{2}^{2})^{2} +  ( 2 n_{2} n_{3}  + 2n_{0} n_{1} )^{2}  }   } \; ] \; ,
\nonumber
\end{eqnarray}

\noindent for $\sin b$ we obtain
\begin{eqnarray}
\sin b = 2 ( n_{0} n_{2} - n_{1}n_{3} ) \; .
\label{4.11}
\end{eqnarray}

Now, let us calculate $\cos b$:
\begin{eqnarray}
\cos b = y_{0}^{2} - y_{2}^{2} =
 {  (n_{0} x_{0} + n_{1} x_{1} )^{2} - (n_{2} x_{0} - n_{3} x_{1})^{2}
\over
(n_{2} x_{0} - n_{3} x_{1})^{2} +  (n_{0} x_{0} + n_{1} x_{1} )^{2}   } \; ,
\nonumber
\end{eqnarray}

\noindent Allowing for expressions for numerator
\begin{eqnarray}
(n_{0} x_{0} + n_{1} x_{1} )^{2} - (n_{2} x_{0} - n_{3} x_{1})^{2}=
\nonumber
\\
= (n_{0}^{2}-n_{2}^{2}){1 + \cos a \over 2} + (n_{1}^{2}-n_{3}^{2}) {1 - \cos a \over 2} + (n_{0}n_{1} +n_{2}n_{3})\sin a
\nonumber
\\
= {1 \over 2} \; [ (n_{0}^{2}-n_{2}^{2}  + n_{1}^{2}-n_{3}^{2}) +
(n_{0}^{2}-n_{2}^{2}  - n_{1}^{2}+ n_{3}^{2}) \cos a +
2 (n_{0}n_{1} +n_{2}n_{3})\sin a \; ] =
\nonumber
\\
= {1 \over 2} \; [ (n_{0}^{2} + n_{1}^{2} -n_{2}^{2}  -n_{3}^{2}) +
\sqrt {
(n_{0}^{2}  + n_{3}^{2}    - n_{1}^{2} -  n_{2}^{2})^{2} +  ( 2 n_{2} n_{3}  + 2n_{0} n_{1} )^{2}  }  \; ]
\end{eqnarray}

\noindent
and  for  denominator
\begin{eqnarray}
(n_{2} x_{0} - n_{3} x_{1})^{2} +  (n_{0} x_{0} + n_{1} x_{1} )^{2} =
\nonumber
\\[3mm]
= { (   n_{0}^{2} + n_{1}^{2} -n_{2}^{2} -n_{3}^{2} )
+  \sqrt {
(n_{0}^{2}  + n_{3}^{2}    - n_{1}^{2} -  n_{2}^{2})^{2} +  ( 2 n_{2} n_{3}  + 2n_{0} n_{1} )^{2}  }
  \over
2\; \sqrt {
(n_{0}^{2}  + n_{3}^{2}    - n_{1}^{2} -  n_{2}^{2})^{2} +  ( 2 n_{2} n_{3}  + 2n_{0} n_{1} )^{2}  }   } \;\; ,
\nonumber
\end{eqnarray}

\noindent
thus $\cos b$ equals
\begin{eqnarray}
\cos b =
\sqrt {
(n_{0}^{2}  + n_{3}^{2}    - n_{1}^{2} -  n_{2}^{2})^{2} +  ( 2 n_{2} n_{3}  + 2n_{0} n_{1} )^{2}  }\; \;.
\label{4.12}
\end{eqnarray}

\noindent It is easily verified identity
\begin{eqnarray}
\sin^{2} b + \cos^{2} b = 4 ( n_{0} n_{2} - n_{1}n_{3} )^{2} + 4(  n_{2} n_{3}  + n_{0} n_{1} )^{2} +
(n_{0}^{2}  + n_{3}^{2}    - n_{1}^{2} -  n_{2}^{2})^{2} =
\nonumber
\\
= 4(n_{0}^{2}  +n_{3}^{2})(n_{1}^{2}+n_{2}^{2}) + (n_{0}^{2}  + n_{3}^{2}    - n_{1}^{2} -  n_{2}^{2})^{2} =
(n_{0}^{2}  + n_{3}^{2}    + n_{1}^{2} +  n_{2}^{2})^{2} = 1 \; .
\nonumber
\end{eqnarray}

In the end of the Section let us collect obtained results:
 the angles $a,b,c$ are determined by relations

\begin{eqnarray}
\cos a = { n_{0}^{2}  + n_{3}^{2}    - n_{1}^{2} -  n_{2}^{2} \over
\sqrt {
(n_{0}^{2}  + n_{3}^{2}    - n_{1}^{2} -  n_{2}^{2})^{2} +  ( 2 n_{2} n_{3}  + 2n_{0} n_{1} )^{2}  } } \; ,
\nonumber
\\
\sin a = {  2  n_{2} n_{3}  + 2 n_{0} n_{1}  \over \sqrt{
(n_{0}^{2}  + n_{3}^{2}    - n_{1}^{2} -  n_{2}^{2} ) ^{2} +  ( 2 n_{2} n_{3}  + 2  n_{0} n_{1} )^{2}  }} \; ,
\nonumber
\end{eqnarray}
\begin{eqnarray}
\cos b =
\sqrt {
(n_{0}^{2}  + n_{3}^{2}    - n_{1}^{2} -  n_{2}^{2})^{2} +  4(  n_{2} n_{3}  + n_{0} n_{1} )^{2}  }=
\nonumber
\\
=
\sqrt{1 - 4( n_{0} n_{2} - n_{1}n_{3} )^{2}} \; , \qquad
\sin b = 2 ( n_{0} n_{2} - n_{1}n_{3} ) \; ,
\nonumber
\end{eqnarray}
\begin{eqnarray}
\cos c = {(n_{0}^{2} - n_{3}^{2} + n_{1}^{2} - n_{2}^{2})\over
\sqrt{  (n_{0}^{2} - n_{3}^{2} + n_{1}^{2} - n_{2}^{2})^{2} + (2n_{0}n_{3}+2n_{1}n_{2})^{2} } } \; ,
\nonumber
\\
\sin c = { 2(n_{0}n_{3}+n_{1}n_{2})  \over
\sqrt{  (n_{0}^{2} - n_{3}^{2} + n_{1}^{2} - n_{2}^{2})^{2} + (2n_{0}n_{3}+2n_{1}n_{2})^{2}  } }\; .
\label{4.12}
\end{eqnarray}

They provide us with solution of the following factorization problem:
\begin{eqnarray}
U = U_{1} U_{2} U_{3} =
 (x_{0} +ix_{1}\sigma_{1})
(y_{0} +iy_{2}\sigma_{2})(z_{0} +iz_{3}\sigma_{3}) \; ;
\nonumber
\\[3mm]
x_{0} = \cos ( a /2) \; , \; x_{1} = \sin (a / 2) \; ,
\nonumber
\\
y_{0} = \cos (b /  2)\; , \; y_{2} = \sin( b / 2) \; ,
\nonumber
\\
z_{0} = \cos ( c / 2)\; , \; z_{3} = \sin (c / 2) \; .
\nonumber
\\
U =  n_{0} + in_{1}\sigma_{1}
+ in_{2}\sigma_{2} + in_{3}\sigma_{3} \; ,
\nonumber
\\
[3mm]
n_{0} = \; \; x_{0} y_{0} z_{0} +  x_{1} y_{2} z_{3} \; ,
\qquad
n_{1} = -x_{0} y_{2} z_{3} +  x_{1} y_{0} z_{0} \; ,
\nonumber
\\
n_{2} = \; \;x_{0} y_{2} z_{0} +  x_{1} y_{0} z_{3} \; ,
\qquad
n_{3} = \;\; x_{0} y_{0} z_{3} -  x_{1} y_{2} z_{0} \; .
\label{4.13}
\end{eqnarray}

\section{All six types of 3-element factorizations}

There exist six different 3-element factorizations:
\begin{eqnarray}
(1) \qquad U_{1}U_{2}U_{3} \; , \qquad \qquad  (1)' \qquad U_{1}U_{3}U_{2} \; ,
\nonumber
\\
(2) \qquad U_{2}U_{3}U_{1} \; , \qquad \qquad  (2)' \qquad U_{2}U_{1}U_{3} \; ,
\nonumber
\\
(3) \qquad U_{3}U_{1}U_{2} \; , \qquad \qquad  (3)' \qquad U_{3}U_{2}U_{1} \; .
\label{5.1}
\end{eqnarray}

Let us compare six  correspondent sets of equations analogues  to (\ref{4.13}):

\begin{eqnarray}
(1)- 123 \; , \qquad n_{0} = \; \; x_{0} y_{0} z_{0} +  x_{1} y_{2} z_{3} \; ,
\qquad
n_{1} = -x_{0} y_{2} z_{3} +  x_{1} y_{0} z_{0} \; ,
\nonumber
\\
n_{2} = \; \;x_{0} y_{2} z_{0} +  x_{1} y_{0} z_{3} \; ,
\qquad
n_{3} = \;\; x_{0} y_{0} z_{3} -  x_{1} y_{2} z_{0} \; ;
\nonumber
\end{eqnarray}

\begin{eqnarray}
(1)' - 132  \; , \qquad
n_{0} = x_{0} z_{0} y_{0} + (- x_{1})  z_{3} y_{2}\; ,  \qquad
-n_{1} =  -x_{0}  z_{3} y_{2} + (- x_{1} )   z_{0} y_{0} \; ,
\nonumber
\\
n_{3} = x_{0}   z_{3} y_{0}  + (- x_{1} )   z_{0} y_{2}  \; , \qquad
n_{2} = x_{0}  z_{0}  y_{2}  - (- x_{1})   z_{3}  y_{0} \; ;
\nonumber
\end{eqnarray}

\begin{eqnarray}
(2) -  231 \; ,\qquad
n_{0} = y_{0} z_{0} x_{0} + y_{2} z_{3} x_{1} \; , \qquad
n_{2} = - y_{0} z_{3} x_{1} + y_{2} z_{0} x_{0}  \; ,
\nonumber
\\
n_{3} =   y_{0} z_{3} x_{0} + y_{2} z_{0} x_{1}
\; , \qquad
n_{1} = y_{0} z_{0} x_{1} - y_{2} z_{3} x_{0} \; ;
\nonumber
\end{eqnarray}

\begin{eqnarray}
(2)' -  213 \; , \qquad
n_{0} = y_{0} x_{0} z_{0} + (- y_{2} )x_{1} z_{3} \; , \qquad
-n_{2}=  - y_{0} x_{1} z_{3} + (- y_{2}) x_{0} z_{0}
 \; ,
\nonumber
\\
n_{1} = y_{0} x_{1} z_{0} + (- y_{2}) x_{0} z_{3}
 \; , \qquad
n_{3} = y_{0} x_{0} z_{3} - (- y_{2}) x_{1} z_{0}
\; ;
\nonumber
\end{eqnarray}

\begin{eqnarray}
(3) - 312  \; ,\qquad
n_{0} = z_{0} x_{0} y_{0} + z_{3} x_{1} y_{2} \; , \qquad
n_{3} = - z_{0} x_{1} y_{2} + z_{3} x_{0} y_{0}\; ,
\nonumber
\\
n_{1} = z_{0} x_{1} y_{0} + z_{3} x_{0}  y_{2}   \; , \qquad
n_{2} = z_{0} x_{0}  y_{2} -  z_{3}  x_{1} y_{0} \; ;
\nonumber
\end{eqnarray}

\begin{eqnarray}
(3)' - 321  \; ,\qquad
n_{0} = z_{0} y_{0} x_{0} + (- z_{3}) y_{2} x_{1} \; , \qquad
-n_{3} = -z_{0} y_{2} x_{1} + (- z_{3} )y_{0} x_{0} \; ,
\nonumber
\\
n_{2} = z_{0} y_{2} x_{0} + (- z_{3}) y_{0} x_{1}
 \; , \qquad
n_{1} = z_{0} y_{0} x_{1} - (- z_{3}) y_{2}  x_{0} \; ;
\label{5.2}
\end{eqnarray}

\noindent
Solutions for all six problems  of (\ref{5.2}) can be obtained from the result (\ref{4.12})  -- (\ref{4.13}) for (123)-variant
with the help of formal changes described in the following  table:

\begin{eqnarray}
\left. \begin{array}{lllll}
  & \qquad \qquad a             & \qquad b             & \qquad c             &                                 \\[3mm]
123  & \qquad \qquad (x_{0},x_{1}) & \qquad (y_{0},y_{2}) & \qquad (z_{0},z_{3}) & \qquad (n_{0},\;\;n_{1},n_{2},n_{3})\\
132  & \qquad \qquad (x_{0},x_{1}) & \qquad  (z_{0},z_{3})& \qquad (y_{0},y_{2}) & \qquad (n_{0},-n_{1},n_{3},n_{2}) \\[3mm]
231  & \qquad \qquad (y_{0},y_{2}) & \qquad (z_{0},z_{3}) & \qquad (x_{0},x_{1}) & \qquad (n_{0},\;\;n_{2},n_{3},n_{1})\\
213  & \qquad \qquad (y_{0},y_{2}) & \qquad  (x_{0},x_{1})& \qquad (z_{0},z_{3}) & \qquad (n_{0},-n_{2},n_{2},n_{3}) \\[3mm]
312  & \qquad \qquad (z_{0},z_{3}) & \qquad (x_{0},x_{1}) & \qquad (y_{0},y_{2}) & \qquad (n_{0},\;\;n_{3},n_{1},n_{2})\\
321  & \qquad \qquad (z_{0},z_{3}) & \qquad  (y_{0},y_{2})& \qquad (x_{0},x_{1}) & \qquad (n_{0},-n_{3},n_{2},n_{2})
     \end{array} \right.
     \label{5.3}
\end{eqnarray}

\section{Polarization of the light in
Mueller-Stokes  and Jones \\  formalisms}

Let  us start with some definitions  concerning the polarization of the light \cite{1992-Snopko}.
For a plane  electromagnetic wave spreading along the axis $z$, in an arbitrary fixed point $z$ we have
\begin{eqnarray}
E^{1} = A \cos \omega t \; , \qquad E^{2} = B \cos (\omega t + \Delta ) \; , \qquad E^{3} = 0
\label{3.1}
\end{eqnarray}

\noindent four Stokes parameters $(S_{a}) =(I, S^{1}, S^{2},S^{3})$ are defined by relations
\begin{eqnarray}
I =\; < E_{1}^{2} + E_{2}^{2} > \; , \qquad S^{3} = \;< E_{1}^{2} - E_{2}^{2} >  \; ,
\nonumber
\\
S^{1} = \;< 2E_{1}E_{2} \; \cos \Delta > \; , \qquad  S^{1} = \;< 2E_{1}E_{2}\;  \sin \Delta > \; ;
\label{3.2}
\end{eqnarray}

\noindent the symbol $ < ...>$  stands for the averaging in time.
If the amplitudes $A,B$ and  the phase shift $\Delta$ do not depend on time in measuring process,
the Stokes parameters equal to
\begin{eqnarray}
S^{0} = I^{(0)} = A^{2} + B^{2} \;  , \qquad  S^{3} = A^{2} - B^{2} \; ,
\nonumber
\\
S^{1} = 2AB\;\cos \Delta  \; , \qquad
S^{2} = 2AB\; \sin \Delta  \; ;
\label{3.3}
\end{eqnarray}

\noindent and an identity
\begin{eqnarray}
S_{a}S^{a} = S_{0}^{2} - S_{j} S_{j} = I^{(0)2} - {\bf S}^{2} = 0 \; , \qquad {\bf S} = I^{(0)} {\bf n} \; .
\label{3.4}
\end{eqnarray}

\noindent holds. In other words, for a steady completely polarized light, the Stoke 4-vector is isotropic.
For a natural (non-polarized)  light, Stokes parameters are trivial:
\begin{eqnarray}
N_{a} = (I_{n} , 0, 0, 0)\; , \qquad  N^{a}  N_{a}  > 0 \; .
\nonumber
\end{eqnarray}

When summing two non-coherent light waves, (1) and (2), their Stokes parameters behave in accordance with the linear law:
\begin{eqnarray}
I_{(1)} +  I_{(2)} \; , \qquad  {\bf S} _{(1)} +  {\bf S}_{(2)}
\nonumber
\end{eqnarray}

A partly polarized light can be obtained when summing two beams of  natural and completely polarized light:
\begin{eqnarray}
N_{a} = (I^{(n)} , 0, 0, 0)\;, \qquad S^{(0)}_{a} = ( I^{(0)},  I^{(0)}  {\bf n} )\; ,
\label{3.5}
\\
S^{a} = N^{a}  + S^{a(0)} = (\; I^{(n)} + I^{(0)} \; )\; (1 , { I^{(0)}  \over  I^{(n)} + I^{(0)}} \; {\bf n} ) \;
\nonumber
\end{eqnarray}

\noindent with notation
\begin{eqnarray}
I = I^{(n)} + I^{(0)} , \qquad p = { I^{(0)}  \over  I^{(n)} + I^{(0)}} , \qquad
\label{3.6}
\end{eqnarray}

\noindent the Stokes vector for a partly polarized light is given in the form
\begin{eqnarray}
S^{a} = I ( 1 , \; p \; {\bf n} ), \qquad   S_{a}S^{a} = I^{2}(1 - p^{2}) \geq 0 \; .
\label{3.7}
\end{eqnarray}

\noindent where  $I$ is a general  intensity, ${\bf n}$ stands for any 3-vector,
$p$ is a degree of polarization which runs within  $[0,\; 1]$ interval:
$0 \leq p \leq 1 $ .

Let us discuss application of SO(3) group theoretical methods to describing  various  polarizing optical devices.

{\bf Pure  polarization attenuators}

($\lambda$-elements changing  uniformly the degree of polarization)

\begin{eqnarray}
\left | \begin{array}{cc}
1 & 0  \\
0 & e^{-\lambda}
\end{array} \right |
 \left | \begin{array}{r}
I \\
I p{\bf n}
\end{array} \right | =  \left | \begin{array}{r}
I' \\
I'  p' {\bf n}'
\end{array} \right | ,   \qquad I' = I ,\;\; {\bf n}' =  {\bf n} \; ,\; \; p' = e^{-\lambda} p \; .
\label{3.7}
\end{eqnarray}

  {\bf  Pure intensity  attenuators}

($\sigma$-elements changing uniformly  the general intensity)

\begin{eqnarray}
\left | \begin{array}{cc}
e^{\sigma}  & 0  \\
0 &  e^{\sigma} I_{3}
\end{array} \right |
 \left | \begin{array}{r}
I \\
I p{\bf n}
\end{array} \right | =  \left | \begin{array}{r}
I' \\
I'  p' {\bf n}'
\end{array} \right | ,   \qquad I' = e^{\sigma} I ,\;\; {\bf n}' =  {\bf n} \; ,\; \; p' =  p \; .
\label{3.8}
\end{eqnarray}

{\bf  Pure polarization rotators }

 SO(3) elements: $R \in SO(3,R)$
\begin{eqnarray}
\left | \begin{array}{cc}
1 & 0  \\
0 & R
\end{array} \right |
 \left | \begin{array}{r}
I \\
I p  {\bf n}
\end{array} \right | =  \left | \begin{array}{r}
I' \\
I' p' {\bf n}'
\end{array} \right | ,   \qquad I' = I ,\;\; {\bf n}' = R {\bf n} \; ,\; \; p' = p \; ,
\label{3.10}
\end{eqnarray}

Now let us consider the Jones formalism and its connection with spinors for rotation and Lorentz groups.
It is convenient to start with a 2-spinor $\psi$, representation of the special linear  group $GL(2.C)$, covering for
the Lorentz group:
\begin{eqnarray}
\psi = \left | \begin{array}{c}
\psi^{1} \\ \psi ^{2}
\end{array} \right |  , \qquad  \Psi ' = B(k) \Psi \; ,
\nonumber
\\
  B(k) = k_{0} + k_{j} \sigma^{j}=
\left | \begin{array}{cc}
k_{0}+k_{3} & k_{1}-ik_{2} \\
 k_{1}+ik_{2} & k_{0}-k_{3}   \end{array} \right | =\left | \begin{array}{cc}
a & d \\
c & b   \end{array} \right |  ,
\nonumber
\\
 \mbox{det} = k_{0}^{2}- {\bf k}^{2} =  ab -cd = 1 \; , \qquad B(k) \in SL(2.C) \; .
\label{4.12}
\end{eqnarray}

From the spinor $\psi$ one may construct a 2-rank spinor $\Psi \otimes \psi^{*}
$, the $2\times 2$ matrix,  which in turn can be resolved in term of
Pauli matrices. we will need two sets: $\sigma^{a} = (I, \sigma^{j})$ and   $\bar{\sigma}^{a} = (I, -\sigma^{j})$.
Let us decompose 2-rank spinor into the sum
\begin{eqnarray}
\Psi \otimes \psi^{*} = {1 \over 2} \;( S_{a} \; \bar{\sigma}^{a} )  =
{1 \over 2} \; (S_{0} - S_{j}\; \sigma^{j} )\; .
\label{4.13}
\end{eqnarray}

The spinor nature of $\psi$ will generate a definite transformation law for $S_{a}$:
\begin{eqnarray}
(\psi' \otimes \psi ^{'*}) = B(k)  (\psi' \otimes \psi ^{'*}) B^{+}(k)  \qquad \Longrightarrow
\qquad
S'_{a} \; \bar{\sigma}^{a} =  B(k)  S_{a} \; \bar{\sigma}^{a} B^{+}(k) \; .
\label{4.14}
\end{eqnarray}

Now, one should use a well-known relation in the theory of the Lorentz group:
\begin{eqnarray}
B(k)   \bar{\sigma}^{a} B^{+}(k) =  \bar{\sigma}^{b}  L_{b}^{\;\;\;\;a}
\label{4.15}
\end{eqnarray}

\noindent where  $L^{\;\;a}_{b}(x)$  is a $4 \times 4$ matrix, defined by
\begin{eqnarray}
L^{\;\;a}_{b}(x) = {1 \over 2} \; sp \; \left [\; \sigma _{b} \; B(k(x))
\bar{\sigma }^{a} \; B(k^{*}(x)) \; \right ] = L^{\;\;a}_{b}
(k(x), \; k^{*}(x)) \;  ,
\label{4.16}
\end{eqnarray}

\noindent With the use of the  known formulas for traces of the Pauli matrices:
\begin{eqnarray}
{1 \over 2 }\; \mbox{sp}\; (\sigma _{k}\bar{\sigma }_{l} \sigma _{a} \bar{\sigma }_{b})
=  g_{kl}  g_{ab} -  g_{ka}  g_{lb}
 + g_{kb} g_{la}   + i \epsilon _{klab} \;  ,
\nonumber
\end{eqnarray}

\noindent for the matrix  $L $  we get
\begin{eqnarray}
L^{\;\; a}_{b} (k,\; k^{*}) = \bar{\delta }^{c}_{b} \; \left [ \;
- \delta^{a}_{c} \; k^{n} \; k^{*}_{n} \;  + \;  k_{c} \; k^{a*}
\; + \; k^{*}_{c} \; k^{a}\; + \; i\; \epsilon ^{\;\;anm}_{c}\;
k_{n} \; k^{*}_{m} \; \right ] \;  ;
\label{4.17}
\end{eqnarray}

\noindent where
\begin{eqnarray}
\bar{\delta }^{c}_{b} =  \left \{ \begin{array}{l}
 0 , \; \;  c  \neq  b \; ; \\
+1 , \; \;  c = b = 0 \; ; \\
-1 , \; \;  c = b = 1, \; 2,\; 3 \; .
\end{array} \right.
\nonumber
\end{eqnarray}

\noindent It should be noted that the matrix $L= (L^{\;\; a}_{b})$ used in Section  transforms contra-variant vector, that is
\begin{eqnarray}
L = L^{a}_{\;\;\;b}\; , \qquad U^{a} =  L^{a}_{\;\;\;b} \; U^{b} , \qquad  L^{a}_{\;\;\;b}  =(L^{-1})_{b}^{\;\;\;b}
\nonumber
\end{eqnarray}

Substituting (\ref{4.15}) into  (\ref{4.14}), one gets the transformation law for   $S_{a}$:
\begin{eqnarray}
S'_{b} = L_{b}^{\;\;\;a} \; S_{a}\; .
\label{4.18}
\end{eqnarray}

\noindent Thus,  spinor transformation $B(k)$ for spinor $\psi$
generates  vector transformation  $L_{b}^{\;\;\;a} (k, k^{*})$.
Different in sign spinor matrices, $\pm B$ lead  to one the same matrix $L$.
If we restrict ourselves to the case of SU(2) group, for matrix $L^{\;\; a}_{b} (k,\; k^{*})$ from  (\ref{4.17})
we get:
\begin{eqnarray}
k_{0}= n_{0} \; , \qquad k_{j} = - i n_{j} \; , \qquad n_{0}^{2}  + n_{j}n_{j} = +1 \; , \qquad
B (n) = n_{0} -i n_{j} \sigma_{j} \; ,
\nonumber
\\[2mm]
L(+n) =L(-n) =
 \left | \begin{array}{rrrr}
1  & 0  &  0  & 0  \\
0 & 1  -2 (n_{2}^{2}+n_{3}^{2})  & -2n_{0}n_{3}+2n_{1}n_{2}  & 2n_{0}n_{2}+ 2n_{1}n_{3}  \\
0 & 2n_{0}n_{3}+2n_{1}n_{2}  &  1  - 2( n_{1}^{2} +n_{3}^{2} )  & - 2n_{0}n_{1} +2n_{2}n_{3}  \\
0 & -2n_{0}n_{2}+2n_{1}n_{3}  &  2n_{0}n_{1}+2n_{2}n_{3}  &  1 -2(
n_{1}^{2}+n_{2}^{2})
\end{array}  \right | \; .
\label{su-2'}
\end{eqnarray}

Let us introduce a polarization  Jones spinor $\psi$:
\begin{eqnarray}
\psi =
\left | \begin{array}{c}
\psi^{1} \\ \psi ^{2}
\end{array} \right | =
\left | \begin{array}{r}
A \;e^{i\alpha}  \\ B\;  e^{i\beta}
\end{array} \right | , \qquad \psi \otimes \psi^{*}=
\left | \begin{array}{cc}
A^{2} & AB e^{-i(\beta-\alpha)} \\
AB e^{+i(\beta-\alpha)} & B^{2}
\end{array} \right |=
\nonumber
\\
= {1 \over 2} (S_{0} - S_{j}\; \sigma^{j}) =
{1 \over 2} \left | \begin{array}{cc}
S_{0} - S_{3} & -S_{1}+ i S_{2}     \\
 -S_{1} - i S_{2}           &  S_{0}+S_{3}
 \end{array} \right | =  {1 \over 2} \left | \begin{array}{cc}
S^{0} + S^{3} & S^{1} - i S^{2}     \\
 S^{1} + i S^{2}           &  S^{0}-S^{3}
 \end{array} \right |\; ,
\label{4.19}
\end{eqnarray}

\noindent
that is
\begin{eqnarray}
A^{2} = {1 \over 2}\; (S^{0} + S^{3})  \; , \qquad   B^{2} = {1 \over 2}\; ( S^{0}-S^{3}) \; ,
\nonumber
\\
 {1 \over 2} \;  (S^{1} + i S^{2} ) = AB e^{i(\beta-\alpha)} \; , \;
{1 \over 2}\; ( S^{1} - i S^{2}  ) = AB e^{-i(\beta-\alpha)} \; .
\label{4.20}
\end{eqnarray}

\noindent
From (\ref{4.20}) it follows
\begin{eqnarray}
   S^{3} = A^{2} - B^{2} \; , \qquad
S^{1} = 2AB \cos (\beta - \alpha)\; , \qquad S^{2} = 2AB \sin (\beta - \alpha)\; ,
\nonumber
 \\
S^{0} = A^{2} + B^{2}  = + \sqrt { S_{1}^{2} +S_{2}^{2} + S_{3}^{2} }\;   .
\label{4.21}
\end{eqnarray}

\noindent
They should be compared with eqs. (\ref{3.3})
\begin{eqnarray}
S^{0} = I^{(0)} = A^{2} + B^{2}  = + \sqrt { S_{1}^{2} +S_{2}^{2} + S_{3}^{2} }\;  ,
\nonumber
 \\
  S^{3} = A^{2} - B^{2} \; , \qquad
S^{1} = 2AB\;\cos \Delta  \; , \qquad
S^{2} = 2AB\; \sin \Delta  \; ;
\label{4.22}
\end{eqnarray}

\noindent
They coincide if $(\beta - \alpha)= \Delta$.
Instead of $\alpha , \beta$ it is convenient introduce new variables:
\begin{eqnarray}
\Delta = \beta - \alpha \;, \qquad  \gamma =  \beta + \alpha \; ,
\label{4.23}
\end{eqnarray}

\noindent correspondingly the spinor $\psi$ will look  (\ref{4.12})
\begin{eqnarray}
\psi =
\left | \begin{array}{c}
\psi^{1} \\ \psi ^{2}
\end{array} \right |  =
e^{i\gamma /2} \;
\left | \begin{array}{r}
A \;e^{- i\Delta /2 }
 \\ B\;  e^{+i\Delta/2}
\end{array} \right |  =
e^{i\gamma /2} \;
\left | \begin{array}{r}
\sqrt{(S+S^{3} )/2}\;  \;e^{- i\Delta /2 }
 \\ \sqrt{(S-S^{3} )/2}\; \;  e^{+i\Delta/2}
\end{array} \right |
.
\label{4.24}
\end{eqnarray}

\noindent Spinor $\psi$  is a Jones complex 2-vector, it is just another representation of the electric
field in the plane electromagnetic wave (see (\ref{3.1})).

\section{Discussion}

Produced formulas describing  all 2-element  and 3-element factorizations
\begin{eqnarray}
U_{2}\;  U_{3} \; U_{2}' \; ,\qquad   U_{1}U_{2}U_{3} \; ,  \qquad \mbox{and so on}
 \nonumber
\end{eqnarray}

\noindent of the elements of the groups SU(2) and SO(3,R) may be
used as a base for corresponding angle parametrization  of the
unitary   group SU(2); also in the context of the light
polarization optics they can  be used as a base to resolve
arbitrary polarization  rotators  into different sets of
elementary ones of two or  tree types:
\begin{eqnarray}
\left | \begin{array}{cc}
1 & 0  \\
0 & R
\end{array} \right |
 \left | \begin{array}{r}
I \\
I p \;  {\bf n}
\end{array} \right | =  \left | \begin{array}{r}
I \\
I p \; {\bf n}'
\end{array} \right | \;  .
\nonumber
\end{eqnarray}

\noindent
The  factorization formulas produced may be  readily extended
to  linear group SL(2.C), spinor covering for Lorentz group $L_{+}^{\uparrow}$.

\section{Acknowledgement}

This  work was  supported  by Fund for Basic Research of Belarus
 F07-314.

\end{document}